\documentclass[doublecol]{epl2}
\usepackage{amsmath}
\usepackage{amssymb}
\usepackage{graphicx}
\usepackage{bm}
\usepackage{dcolumn}
\usepackage{subfigure}
\usepackage{psfrag}
\usepackage{float}
\usepackage{subeqnarray}
\usepackage{ltxgrid}
\usepackage{lipsum}

\title{Position-dependent effect of non-magnetic impurities on superconducting properties of nanowires}
\shorttitle{Effect of impurities on superconducting nanowires} %Insert here a short version of the title if it exceeds 70 characters

\author{L.-F. Zhang\and L. Covaci\and F. M. Peeters}
\shortauthor{L.-F. Zhang \etal}

\institute{
   Departement Fysica, Universiteit Antwerpen,
Groenenborgerlaan 171, B-2020 Antwerpen, Belgium }

\abstract{Anderson's theorem states that non-magnetic impurities do not change the bulk properties
of conventional superconductors.  However, as the dimensionality is reduced, the effect of
impurities becomes more significant. Here we investigate superconducting nanowires with diameter
{comparable to the Fermi wavelength $\lambda_F$ (which is less than the superconducting
coherence length)} by using a microscopic description based on the Bogoliubov-de Gennes method.  We
find that: 1) impurities strongly affect the superconducting properties, 2) the effect is impurity
position-dependent, and 3) it exhibits opposite behavior for resonant and off-resonant wire widths.
We show that this is due to the interplay between the shape resonances of the order parameter and
the sub-band energy spectrum induced by the lateral quantum confinement.  These effects can be used
to manipulate the Josephson current, filter electrons by subband and investigate the symmetries of
the superconducting subband gaps.}

\begin{document}

\maketitle
\twocolumngrid
\section{Introduction}

The effect of impurities on superconductivity have intrigued scientists for several decades.  Adding impurities to a superconductor can not only provide an useful tool to investigate the superconducting state but can also be used to optimize the performance of certain devices.  For example, impurities are used to distinguish between various symmetries of the superconducting state\cite{revmod1, probe}, to identify topological superconductors\cite{HuiHu} or to pin vortices in order to enhance the superconducting critical current\cite{pin}.  Strong disorder is often used to study the superconducting-insulator transitions and the localization of the Cooper pairs\cite{disorder1, disorder2, disorder3, disorder4, disorder5, disorder6}.  It is well known that magnetic impurities, which are pair breakers, suppress superconductivity. On the other hand, as Anderson's theorem points out\cite{Anderson}, for bulk conventional superconductors, small concentrations of non-magnetic impurities do not change the thermodynamic properties of the system, such as the superconducting critical temperature $T_c$, and that impurities have only local effects\cite{revmod1} on the superconducting order parameter (OP) and the density of states.

On the other hand, quasi-one-dimensional superconducting nanowires show different properties from
the bulk and could have wide potential applications in the near future\cite{nanoscale}.
{First, if in some regions, the diameter of the cross section of the wire is shorter than
the superconducting coherence length, $\xi$ (also known as the healing length of the OP),
superconductivity weakens there, thus affecting transport properties of the wires.}  Quantum phase
slips, leading to the loss of phase coherence, are one of the main reasons for the appearance of
normal regions giving rise to a finite resistance below $T_c$ in nanowires\cite{slip1, slip2, slip3,
slip4, slip5, slip6, slip7}.  Phase slip junctions can be fabricated in order to take advantage of
this mechanism\cite{slip8}.  Normal regions can also be caused by interactions with photons and this
can be used to build single-photon detectors\cite{app4, app5}.  Second,  surface effects in
nanoscale superconductors result in topological superconducting states\cite{topo}.  Majorana
fermions, which are their own anti-particles, were predicted as quasi-particles in such p-wave
nanowires\cite{majorana1, majorana2}.  Third, quantum confinement results in quantum size
effects\cite{qse1, qse2, qse3}, quantum-size cascades\cite{cascades}, facilitates the appearance of
new Andreev-states\cite{andreev1} and give rise to unconventional vortex states\cite{zhang} induced
by the wavelike inhomogeneous spatial distribution of the OP\cite{qse2}.

{Up to now, the theoretical study of superconducting nanowires was limited to defect-free
cases in the clean limit or to systems with weak links, which were treated in the one-dimensional
limit.}  However,
experiments such as Scanning Tunneling Microscopy (STM) are nowadays able to measure the local
density of states (LDOS) with atomic resolution\cite{STM0,STM1,STM2,STM3,STM4} and consequently the
finite width of nanowires can no longer be neglected. It is known that in low dimensional
superconductors phase fluctuations should play an important role \cite{phase1,phase2,phase3}. The
calculation presented here is done at zero temperature and at mean-field level, therefore it cannot
describe fluctuations. Nevertheless, by calculating the modifications induced by the impurity on the
Josephson current we are able to infer the phase robustness of the condensate, therefore indirectly
provide information about the enhancement of fluctuations and the appearance of phase slips.

In this Letter, we investigate the effect of a nonmagnetic impurity on superconducting nanowires {with diameter comparable to the Fermi wavelength, $\lambda_F$.  The bulk coherence length at zero temperature, $\xi_0$, is much larger than $\lambda_F$.} For such systems, quantum confinement effects dominate and a description based on the
microscopic Bogoliubov-de Gennes (BdG) equations is required.  The impurity, which induces potential
scattering on single electrons, not only affects local properties such as the OP and the LDOS, but
also has global effects on the critical supercurrent.  Quantum confinement leads to inhomogeneous
superconductivity and therefore the effect of the impurity strongly depends on its transverse
location.

The letter is organized as follows: first we briefly present our numerical method and the properties
of the nanowires in the clean limit, then we show the effect of non-magnetic impurities on the
profile of the order parameter and the local density of states. At the end we investigate the
impurity effect on the Josephson critical current.

\section{Theoretical approach}
We start from the well-known BdG equations:
\begin{eqnarray}
\label{BdG 1}  \left[K_0-E_F\right] u_n(\overrightarrow{r})+\Delta(\overrightarrow{r})v_n(\overrightarrow{r})&=&E_nu_n(\overrightarrow{r}), \\
\label{BdG 2}  \Delta(\overrightarrow{r})^\ast u_n(\overrightarrow{r})-\left[K_0^\ast-E_F\right]
v_n(\overrightarrow{r})&=&E_nv_n(\overrightarrow{r}),
\end{eqnarray}
where $K_0=-(\hbar\nabla)^2/2m+U(\overrightarrow{r})$ is the kinetic energy with $U$ being the potential barrier induced by the impurity and $E_F$ the Fermi energy, $u_n$($v_n$) are electron(hole)-like quasiparticle eigen-wavefunctions, $E_n$ are the quasiparticle eigen-energies.  The impurity potential is modeled by a symmetrical Gaussian function, $U(\overrightarrow{r})=U_0exp[-(\overrightarrow{r}-\overrightarrow{r_0})^2/2\sigma^2]$
where $U_0$ is the amplitude, $\overrightarrow{r_0}$ is the location of the impurity and $\sigma$ is the width of the Gaussian.

The pair potential is determined self-consistently from
the eigen-wavefunctions and eigen-energies:
\begin{equation}\label{OP}
\Delta(\overrightarrow{r})=g\sum\limits_{E_n<E_c}u_n(\overrightarrow{r}) v^\ast_n(\overrightarrow{r})[1-2f(E_n)],
\end{equation}
where $g$ is the coupling constant, $E_c$ is the cutoff energy, and $f(E_n)=[1+\exp(E_n/k_BT)]^{-1}$ is the Fermi distribution function, where $T$ is the temperature.
The local density of states is calculated as usual:
$N(\overrightarrow{r},E)=\sum\limits_{n} [\delta(E_n-E)|u_n(\overrightarrow{r})|^2+\delta(E_n+E)|v_n(\overrightarrow{r})|^2].$

For simplicity, we consider a two-dimensional problem and introduce a long unit cell whose area is
$S=L_xL_y$ where $L_x$ is the length of the unit cell and $L_y$ is the wire width.  Periodic
boundary conditions are set in the $x$ direction and $L_x$ is set to be long enough such that
physical properties are $L_x$-independent.  Due to quantum confinement in the transverse direction
$y$, we set Dirichlet boundary condition at the surface (i.e.
$u_n(\overrightarrow{r})=v_n(\overrightarrow{r})=0,\; r \in \partial S$).

In order to solve more efficiently the self-consistent BdG equations (\ref{BdG 1})-(\ref{OP}), we expand $u_n$($v_n$) by using a two step procedure.  First, in order to get the eigenstates of the single-electron Schr\"odinger equation $K_0\phi_l=E_l\phi_l$ we Fourier expand $\phi_l$:
\begin{equation}\label{Fourier}
\phi_l(x,y) =\sqrt{\frac{2}{L_xL_y}}\sum_{j>0,k} c_{j,k}\exp\left(i k x\right)\sin\left(\frac{\pi j y}{L_y}\right),
\end{equation}
where $c_{j,k}$ are the coefficients of the expansion and wave vector $k=2\pi m/L_x,~m \in \mathbb{Z}$.  Here we use a large number of basis functions in order to ensure the accuracy of the results.  Next, we expand $u_n$($v_n$) in terms of $\phi_l(x,y)$.  In this step, only states with energies $E_l<E_F+\varepsilon$ are included.  In our calculation, $\varepsilon$ is taken to be $30E_c$ and we find that any larger cut-off does not modify the results. We have checked our results, for a specific choice of parameters,  against a more computationally intensive finite-difference method approach.

We choose to study $NbSe_2$ nanowires as an example because of the availability of high-quality
nanowires with $2-25$nm in diameter \cite{nbse1,nbse2}.  Also, theoretical calculations are in good
agreement with experiments, especially, for vortex lines\cite{vortex1, vortex2, vortex3}.  In
addition, the lower Fermi energy makes the calculations feasible.  The parameters of bulk $NbSe_2$
are the following: $m=2m_e$, $E_F=40meV$, $E_c=3meV$ and coupling constant $g$ is set so that the
bulk gap $\Delta_0=1.2meV$, which yields $T_c\approx 8.22K$, $\xi_0=14.7nm$ and $k_F\xi_0=21.23$.
In nanowires, the mean electron density $n_e$ is kept to the value obtained when $L_x,L_y
\rightarrow \infty$ by using an effective $E_F$, where $n_e=\frac{2}{S}\sum_{n} \int \left \{
|u_n|^2f(E_n)+|v_n|^2[1-f(E_n)]\right \}$.  All the calculations are performed at zero temperature.

\section{Numerical Results}

%
%
%\begin{center}
\begin{figure}[ttt]
\includegraphics[width=\columnwidth]{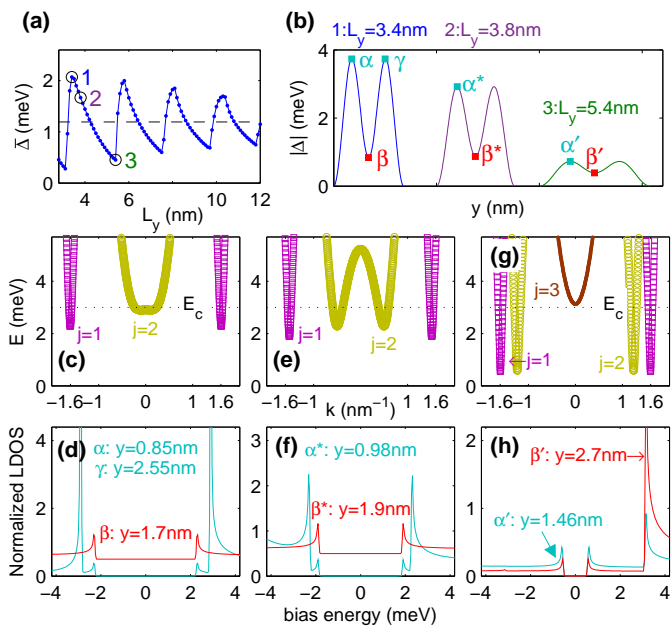}
\caption{ (Color online) Properties of clean $NbSe_2$ nanowires.  (a) Spatially averaged
$\bar{\Delta}$ as a function of $L_y$.  {The open circles $1-3$ indicate the resonance case
($L_y=3.4nm$), intermediate case ($L_y=3.8nm$) and off-resonance case ($L_y=5.4nm$), respectively.}  (b) OP $|\Delta(y)|$ for the three cases.  The $\alpha\beta\gamma$ are defined as the positions where the OP has a local maximum $/$ minimum for resonance, the $\alpha^*\beta^*$ are for intermediate and the $\alpha'\beta'$ are for off-resonance.  The value of these positions are shown in the inset of the panels (d),(f) and (h), respectively.  (c) and (d) energy
spectrum and the corresponding LDOS at positions $\alpha$, $\beta$ and $\gamma$ for resonance. {(e) and (f) the same but for the intermediate case and positions $\alpha^*$ and $\beta^*$.} (g) and (h) the same but for off-resonance and positions $\alpha'$ and $\beta'$.  Note that in panel (d) and (f) the LDOS for position $\beta$ and $\beta^*$ are shifted for clarity. } \label{fig.1}
\end{figure}
%\end{center}
%

% Results
We first show in Fig.~\ref{fig.1} important superconducting properties of clean nanowires, i.e.
with $U\equiv 0$.  The spatially averaged OP, $\bar{\Delta}$, shows quantum size
oscillations as a function of the width, $L_y$, as seen from Fig.~\ref{fig.1}(a).  The resonant
enhancements appear almost regularly with a period of half the Fermi wavelength, i.e. $\lambda_F/2$.
 They are due to the fact that the bottom of the relevant single-electron subbands passes through
the Fermi surface, which results in a significant increase in the density of states, i.e. in the
number of electrons which can form Cooper-pairs.  The different energy spectra for the resonance and
off-resonance cases are the key to understand the behavior of the nanowires.  {For
completeness, we show in Fig.~\ref{fig.1}(b)-(h) more information for the resonance case, with
$L_y=3.4nm$, intermediate case, with $L_y=3.8nm$, and off-resonance case, with
$L_y=5.4nm$\cite{cases}.}  The energy spectrum for resonance [see Fig.~\ref{fig.1}(c)] shows two
energy gaps: a smaller gap for subband $j=1$ and a larger one for $j=2$.  Most quasiparticle states
are just below the cutoff energy $E_c$ for subband $j=2$.  This results in an enhancement of the OP,
which shows two pronounced peaks and is strongly inhomogeneous in the $y$ direction [see
Fig.~\ref{fig.1}(b)].  We consider the spatial positions $\alpha\beta\gamma$ and $\alpha'\beta'$ as
given in the insets of Fig.~\ref{fig.1}(d)(h) where the OP has either a maximum or a minimum. The
corresponding LDOS at positions $\alpha$ and $\beta$ are shown in Fig.~\ref{fig.1}(d).  We notice
that there are two gaps at $\alpha$ and only one gap at $\beta$.  The smaller sub-gap at $\alpha$
and the gap at $\beta$ comes from the $j=1$ subband.  The main gap at $\alpha$ comes mostly from the
$j=2$ subband which is indicated by the ratio of the LDOS peaks at the main gap and at the sub-gap.
In contrast, the off-resonance case shows the same energy gap for $j=1$ and $j=2$ [see
Fig.~\ref{fig.1}(g)] and the LDOS [see Fig.~\ref{fig.1}(h)] is more conventional: only one
superconducting gap and the amplitude of the LDOS is proportional to $|\Delta(y)|$, i.e. the LDOS at
$\alpha'$ (local maximum) is larger than the one at $\beta'$ (local minimum).  Besides the sub-gap
shift seen in the energy spectrum, another important difference between the two cases is that for
resonance
there is always a large number of low momentum quasi-particles [see Fig.~\ref{fig.1}(c)] which are
involved in pairing, whereas there are no quasi-particles at $k=0$ in the off-resonance case [see
Fig.~\ref{fig.1}(e)].  Note that subband $j=3$ in off-resonance sits just outside of $E_c$ and
generates a large asymmetric peak in the LDOS [see Fig.~\ref{fig.1}(f)].  {The intermediate
case [see Fig.~\ref{fig.1}(e) and (f)] mixes the characteristic from both resonance and
off-resonance
cases where it has the intermediate gap difference between subbands $j=1$ and $j=2$ and the
intermediate $k$-value for most quasi-particles states with subband $j=2$.}

%
%\begin{center}
\begin{figure}
\includegraphics[width=\columnwidth]{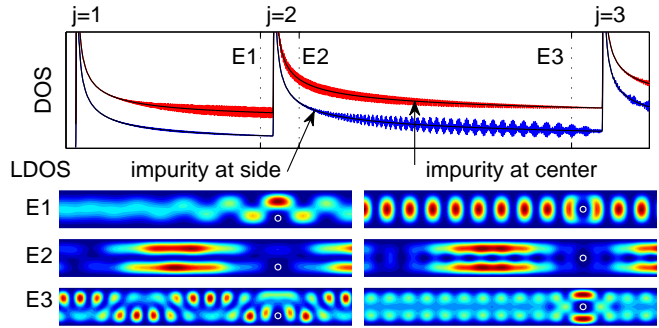}
\caption{ {(Color online) Effect of the impurities on the electronic structure when the nanowire is in normal state.  (upper panel) DOS as a function of energy for an impurity at sided position (blue solid line) and at centered position (red solid line, shifted up for clarity), respectively.  The black solid lines are for $U=0$ as references.  (lower panels) Schematic corresponding LDOS($x,y$) at energy $E1$, $E2$ and $E3$, respectively.  The open circles indicate impurities' position.} } \label{fig.12}
\end{figure}
%\end{center}
%

%
%\begin{center}
\begin{figure}[ttt]
\includegraphics[width=\columnwidth]{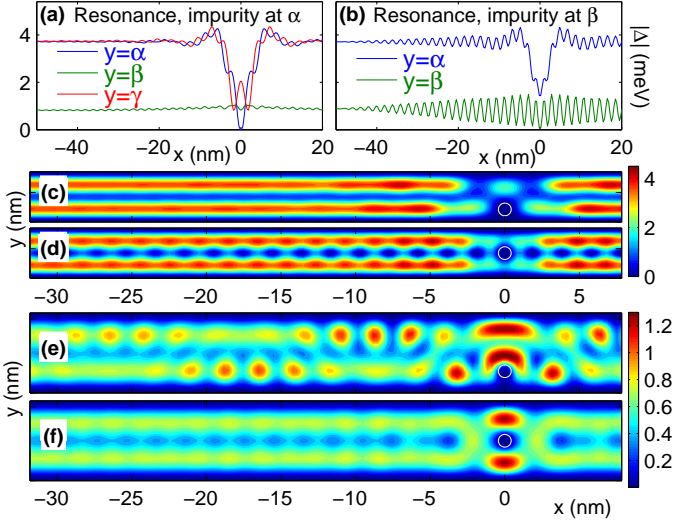}
\caption{ (Color online) OP in the presence of an impurity.  (a) and (c) show $|\Delta(x)|$ at y=$\alpha$, $\beta$, $\gamma$ (defined in Fig.~\ref{fig.1}) and contour plot of $|\Delta(x,y)|$ for an impurity sitting at $\alpha$ in resonance case, respectively.  (b) and (d) are the same as (a) and (c) but for the impurity sitting at the center $\beta$. (e) and (f) show contour plot of $|\Delta(x,y)|$ for an impurity sitting at $\alpha'$ and $\beta'$ in off-resonance case, respectively.  The white open circles indicate impurities' profile where $U=0.1U_0$.} \label{fig.2}
\end{figure}
%\end{center}
%

%
%\begin{center}
\begin{figure}[ttt]
\includegraphics[width=\columnwidth]{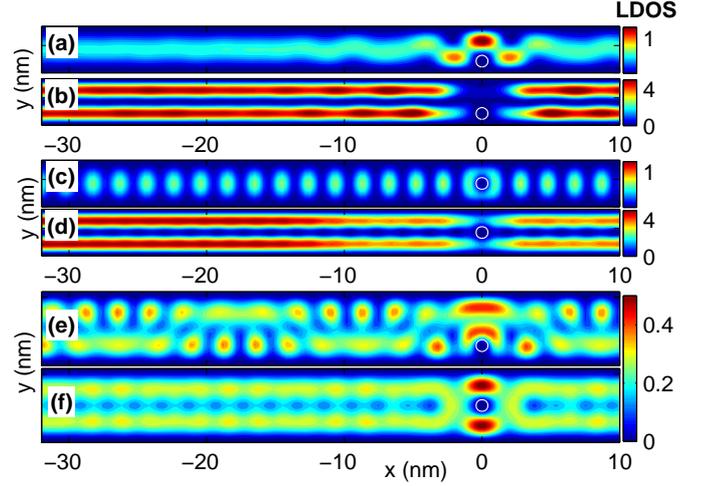}
\caption{ (Color online) Contour plots of LDOS at the energies' (sub-)gap in the presence of an impurity.  (a-d) are for the resonance case with an impurity sitting at $\alpha$ at sub-gap (a) $E/meV= 2.32$ and gap (b) $E/meV=2.93$ and with an impurity sitting at $\beta$ at sub-gap (c) $E/meV=2.36$ and gap (d) $E/meV=2.93$.  (e) and (f) are for off-resonance case with the impurity at $\alpha'$ and at $\beta'$ at gap $E/meV=0.68$ and $0.65$, respectively.  The open circles indicate impurities' profile where $U=0.1U_0$.} \label{fig.3}
\end{figure}
%\end{center}
%
%
%

{Next, we consider an impurity in the nanowire.  The impurity is strong enough to suppress
completely the wave-functions of the electronic states locally.  Meanwhile, the spread of the
impurity is smaller than the width of the nanowire.  The effects of the impurity on the electronic
structures of normal state are presented in Fig.~\ref{fig.12}.  The scattering due to the single
impurity does not change the quasi one-dimensional characteristic seen in the DOS.  However, it
results in additional small oscillations when comparing with the clean case with $U=0$.  In
addition, the oscillations show different pattern for impurity at sided position and at centered
position.  It indicates the influence of the impurity depending on its position and the quantum
number of the electronic state, i.e. $j$ and $k$.  Due to the sine-shaped transverse electronic
wave-functions, the impurity at center position affects the states with subband $j=1$ more than
states with subband $j=2$.  Thus, the DOS of the impurity at center position shows stronger
oscillations than the one of the impurity at side position between the peaks of $j=1$ and $j=2$.
For the same reason, it shows opposite behavior in DOS between the peak of $j=2$ and $j=3$.
Furthermore, the impurity leads to fast oscillations in the LDOS at energies $E1$ and $E3$ [see
Fig.~\ref{fig.12} (lower panels)] because these states have large wave-number, i.e. $k$.  In
contrast, the impurity leads to oscillations with long wave-length in the LDOS at energy $E2$, due
to the states with small $k$-value dominating near $E2$.}

Now we move on to the superconducting state and consider an impurity at position $(0,\alpha)$ and $(0,\beta)$ for resonance case and at $(0,\alpha')$ and $(0,\beta')$ when off-resonance.  We set $U_0=20E_F$ and $\sigma=0.02nm$ so that the impurity strongly suppresses the local OP and the spread of the potential (full width at $1/10$th of maximum) is $0.86nm$, which is shorter than the width of the nanowire.  We show in Fig.~\ref{fig.2} the profile of the amplitude of the OP in the presence of the impurity.  For the resonance, [see Figs.~\ref{fig.2}(c)(d)], the impurity suppresses the OP over the whole width which is more pronounced when in $\alpha$ than in $\beta$.  Meanwhile, the asymmetrical impurity, at $\alpha$, results in a local enhancement at position $(0,\gamma)$, as seen from Fig.~\ref{fig.2}(a).  For the impurity sitting in the center, $\beta$, the OP oscillates along the wire over a longer distance although the impurity sits at a local minima of the OP which intuitively should have a smaller effect.  As seen from Fig.~\ref{fig.2}(b), the OP oscillations extend to $x=\pm 50nm$, which is much farther than the extent of the oscillations for the impurity at $\alpha$.  The reason is that the centered $\beta$ impurity strongly affects the $j=1$ subband quasi-particles due to the sine-shaped transverse wave-function, leaving the $j=2$ subband unaffected.  Moreover, only high $k$ quasi-particles in subband $j=1$ play a role so that scattering on the impurity results in oscillations of the wave-functions.  On the other hand, the impurity at $\alpha$ affects mostly subband $j=2$  but all states with $j=2$ have low $k$ so that the OP recovers fast and has longer wave-length oscillations.  The local density of states in Figs.~\ref{fig.3}(a-d) show evidence that supports this explanation.
For the off-resonance case [see Figs. \ref{fig.2} (e-f)], due to the combination of the $j=1$ and $j=2$ subbands which now have only high $k$ quasi-particles, the impurity always induces strong oscillations in the OP.  Note that the OP is affected more strongly when the impurity sits at a local maximum.  The LDOS shown in Figs.~\ref{fig.3} (e-f) show the same patterns as the OP.

Finally, we study the effect of the impurity on the transport properties such as the Josephson current.  In order to do so, we set a junction link of length $L_j=30nm$ at the center of the nanowire where $\Delta$ is obtained self-consistently.  Outside the link, we fix the phase of the order parameter and impose a phase difference $\delta\theta$ between the two sides of the link, i.e. $\Delta(x<-15nm)=|\Delta|e^{i0}$ and $\Delta(x>15nm)=|\Delta|e^{i\theta}$.  Then, the supercurrent induced by the phase difference $\delta\theta$ can be calculated as follows:
\begin{align*}\label{current}
\vec{J}(\vec{r}) &= \frac{e\hbar}{2mi} \sum_{E_n<E_c}\left \{ f(E_n)u_n^*(\vec{r})\bigtriangledown u_n(\vec{r}) \right.\\
&+ \left. [(1-f(E_n)]v_n(\vec{r})\bigtriangledown v_n^*(\vec{r})-h.c.\right \}.
\end{align*}
Please note that $\vec{J}$ satisfies the continuity condition $\nabla \cdot \vec{J} = 0$ in the link due to the self-consistent $\Delta$\cite{strinati,covaci2006}. Outside the link, $\vec{J}$ is discontinuous due to the fixed phase of the order parameter but these areas can be treated as current sources.

%
%
%
%\begin{center}
\begin{figure}[ttt]
\includegraphics[width=\columnwidth]{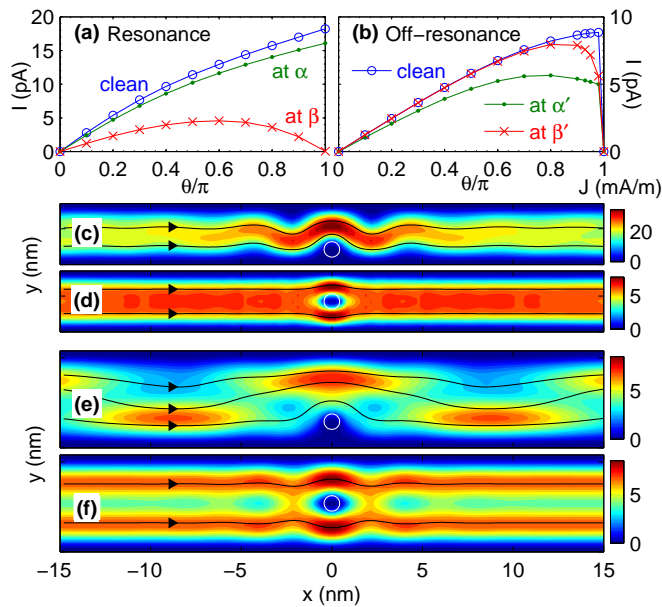}
\caption{ (Color online) Effects of an impurity on Josephson current where the junction length is $L_j=30nm$.  (a) Current-phase relation in the resonance case for clean limit, with impurity at $\alpha$ and at $\beta$.  (b) Results in the off-resonance case for clean limit, impurity at $\alpha'$ and at $\beta'$.  (c)-(f) The spatial distributions of current for an impurity at $\alpha$, $\beta$ for resonance case and at $\alpha'$, $\beta'$ for off-resonance case with phase differences where currents reach their critical Josephson currents.  The color indicates the amplitude of the current and the streamlines indicate the direction of the current. The open circles indicate the impurity profile for which $U=0.1U_0$.}\label{fig.4}
\end{figure}
%\end{center}
%
Figs.~\ref{fig.4}(a) and (b) show the $J-\theta$ relation for resonant and off-resonant cases, respectively.  For resonance, the current increases monotonically with the phase for the impurity at $\alpha$ but it shows a sine-shaped curve for the impurity at $\beta$.  We find that the critical Josephson current is suppressed dramatically for the impurity at $\beta$ while the impurity at position $\alpha$ has little effect.  The explanation is that when the current is small, only the lowest quasi-particle states are involved.  and almost all such states are from the $j=1$ branch and as a result the current distribution $j(y) \propto \sin(\pi y/L_y)$ is as in the clean limit.  This can be seen from Fig.~\ref{fig.4}(d) for $x$ far away from the impurity.  Thus, the impurity blocking effect on the current for the impurity at the center $\beta$ has a larger effect than the off-center position $\alpha$.  In contrast, for the off resonance case, the current with the impurity at $\alpha'$ is more suppressed.  The reason is the same but, here, the $j=1$ and $j=2$ branches contribute both to the current, and as a consequence the current distribution is $j(y) \propto |\sin(2\pi y/L_y)|+|\sin(\pi y/L_y)|$ and has a maximum at the $\alpha'$ position. This can also be seen from Figs.~\ref{fig.4}(f) far away from the impurity. This position dependent effect could be used to investigate the nature of the superconducting condensate, e.g. by using Scanning Gate Microscopy(SGM) in order to locally mimic the effect of the impurity and suppress the order parameter. By monitoring the change in the Josephson current, one can map out the symmetry of the order parameter and the distribution of the supercurrent along the $y$ direction.

\section{Conclusions}

{In conclusion, we studied the effect of non-magnetic impurities on narrow superconducting nanowires in the clean limit, in which quantum confinement plays an important role. By applying BdG theory to the case of $NbSe_2$,} we uncovered several regimes in which the impurity affects the superconducting properties of the nanowire in different
ways. First, depending whether the nanowire is in the resonant or off-resonant regime, the OP will
show slow or fast oscillations away from the impurity, respectively. This is due to the different
nature of the quasi-particles involved in the formation of the Cooper pairs, i.e. small or large
momentum. Additionally, the impurity has a strong position-dependent effect on the Josephson
critical current with opposite behavior in the resonant and off-resonance cases. In the resonant
case an impurity at the center of the wire will strongly suppress the current while for the
off-resonance case the current is slightly suppressed for an impurity sitting at an off-center
location.

{In experiments, the crystal structure of the material, the surface roughness of the
specimen and the properties of the substrate will modify the Fermi level, the band structure, the
electronic wave functions or the electron mean free path.  All these factors could broaden the
single-electron levels and modify the specific scattering patterns of the OP and electronic
structures (LDOS).  However, the superconducting shape resonances are robust, as shown in
more realistic theoretical models\cite{substrate2014}.  In this case, quantum confinement dominates
and the position dependence of the amplitude of the electronic wave functions and OP are robust.  As
a result, our conclusions about the position-dependent impurity effect (especially the effect
on the Josephson current) will not be significantly altered, although specific details might
change.}

We believe that these effects could be used to investigate the nature of the
superconducting condensate and the scattering of the various subbands on the impurity. Also in
realistic nanowires, which contain impurities, one should see a strong impurity effect in the
resonant case. Although the mean-field calculation presented here cannot describe fluctuations, by
showing that the Josephson current can be strongly suppressed in the resonant case, one can infer
that fluctuations will become more important. Our calculations could also provide a basis for a
phenomenological toy model of a 1D disordered Josephson array with position impurity
dependent junction parameters.

\acknowledgments

This work was supported by the Flemish Science Foundation (FWO-Vlaanderen) and the Methusalem funding of the Flemish Government.

\pagebreak

\renewcommand\thefigure{S\arabic{figure}}
\setcounter{figure}{0}
%\begin{widetext}
\onecolumngrid
\begin{center}
\large{\bf Supporting Online Materials}
\end{center}
\vspace{1cm}
\twocolumngrid
%\end{widetext}
%

We present results for $NbSe_2$ nanowires for the resonance case $(L_y=5.8nm)$ and off-resonance case $(L_y=7.5nm)$, where there are three peaks in $\Delta(y)$ now.

\begin{figure*}
\includegraphics[width=1.4\columnwidth]{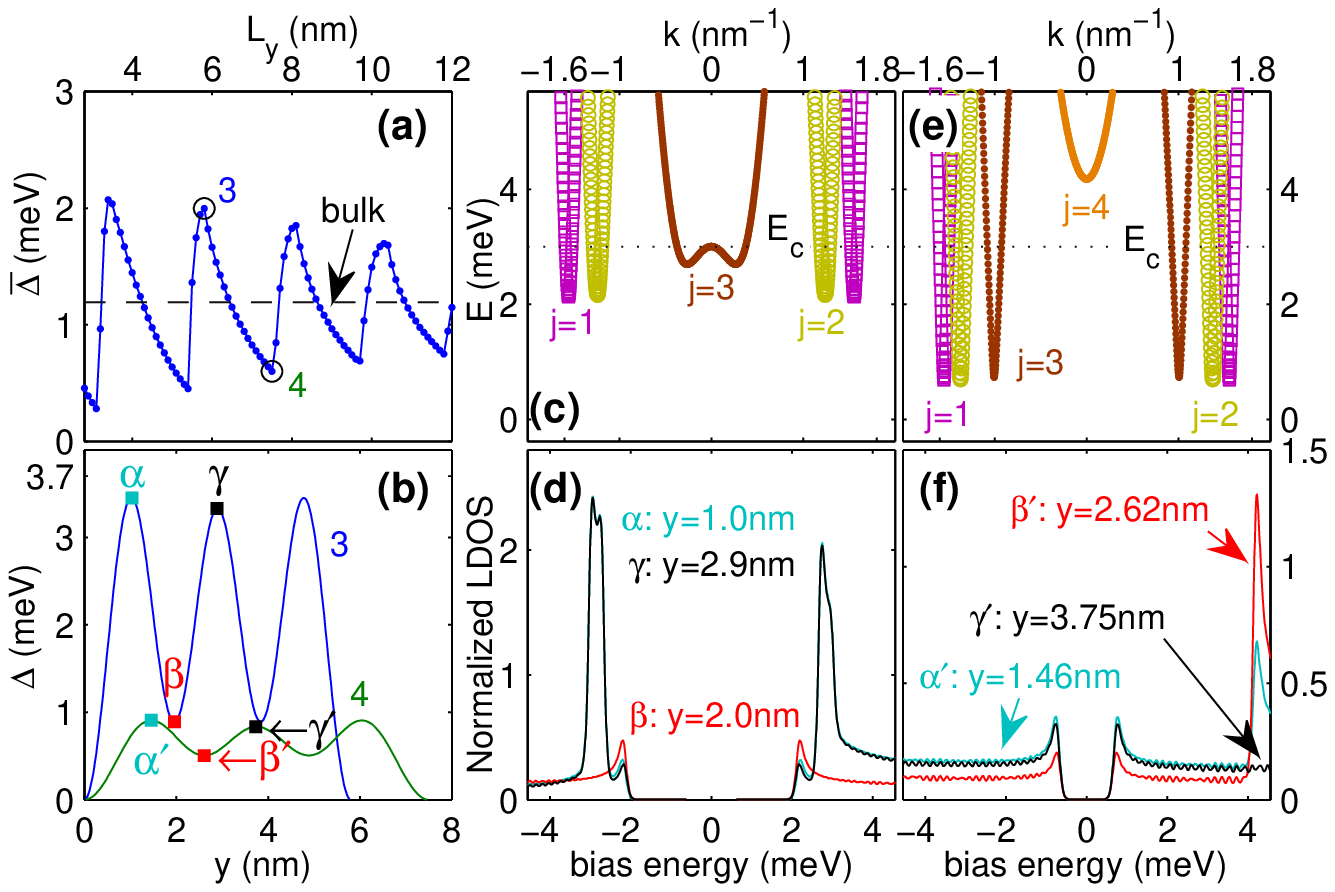}
\caption{ Properties of the clean $NbSe_2$ nanowires.  (a) Spatially averaged $\bar{\Delta}$ as a function of $L_y$.  The open circles $3$ and $4$ indicate the resonance case ($L_y=5.8nm$) and off-resonance case ($L_y=7.5nm$), respectively.  (b) Order parameter $|\Delta(y)|$ for resonance and off-resonance cases.  The $\alpha\beta\gamma$ are defined as the positions where the order parameter has a local maximum $/$ minimum for resonance and the $\alpha'\beta'\gamma'$ for off-resonance.  The value of these positions are shown in the inset of the panels (d) and (f).  (c) and (d) energy spectrum and the corresponding LDOS at positions $\alpha$, $\beta$ and $\gamma$ for resonance.  (e) and (f) the same but for off-resonance and positions $\alpha'$, $\beta'$ and $\gamma'$.} \label{sfig.1}
\end{figure*}

First, we show in Fig.~\ref{sfig.1} the electronic properties in the clean limit for both cases.  They are similar to the results we presented in the main paper.  The difference is that, for the resonance case here, the energy spectrum shows a subgap for both subbands $j=1$ and $j=2$ and the main gap for the subband $j=3$.  Only the quasi-particles of the subband $j=3$ have low momentum $k\approx0$.  Note that the folding of the subband $j=3$ around $k=0$ results in a double peak in the LDOS [see Fig.~\ref{sfig.1}(d)].  For the off-resonance case, all three subbands mix and have only one gap.  Furthermore, no quasiparticle state has low momentum.

\begin{figure*}
\includegraphics[width=1.4\columnwidth]{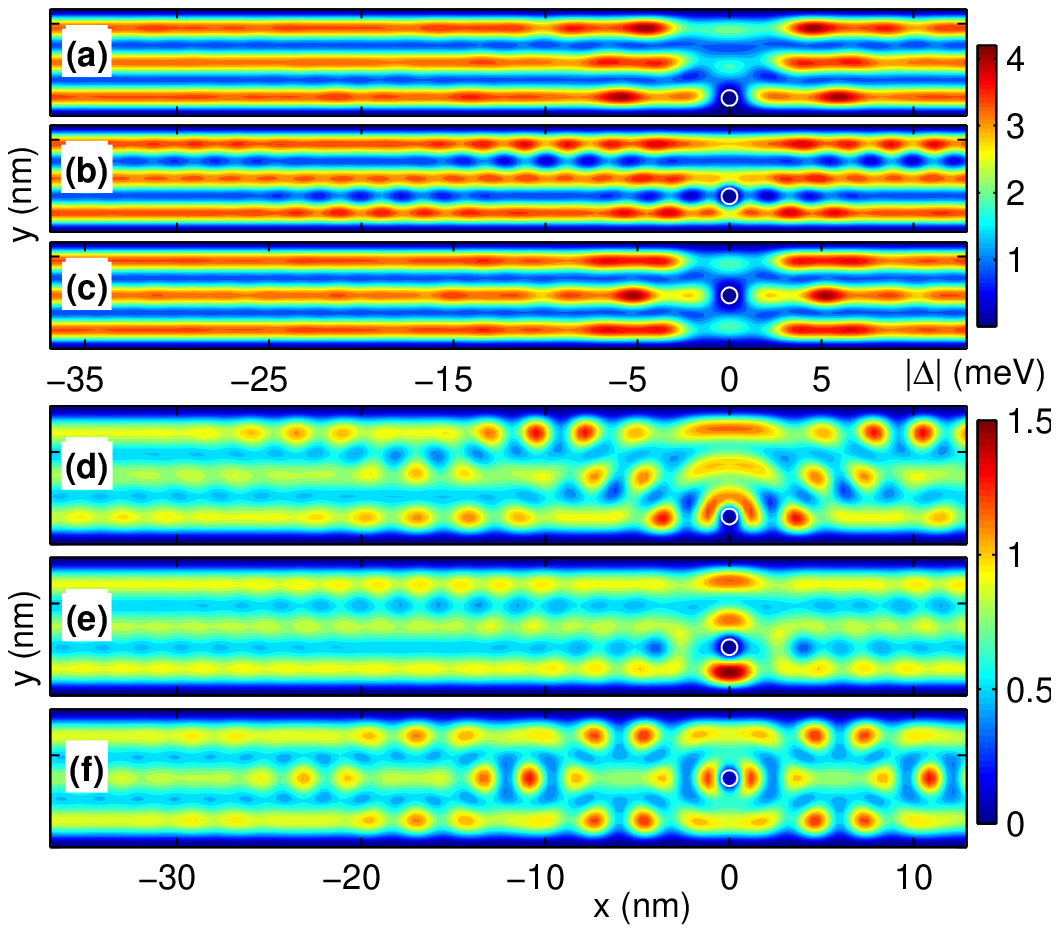}
\caption{ Contour plots of the order parameter in the presence of an impurity.  (a-c) $|\Delta(x,y)|$ for an impurity  sitting at $\alpha$, $\beta$ and $\gamma$ (defined in Fig.~\ref{sfig.1}) in resonance case.  (d-f) are the same as (a-c) but for the impurity sitting at $\alpha'$, $\beta'$ and $\gamma'$ in resonance case.  The white open circles indicate impurities' profile where $U=0.1U_0$.} \label{sfig.2}
\end{figure*}

\begin{figure*}
\includegraphics[width=1.4\columnwidth]{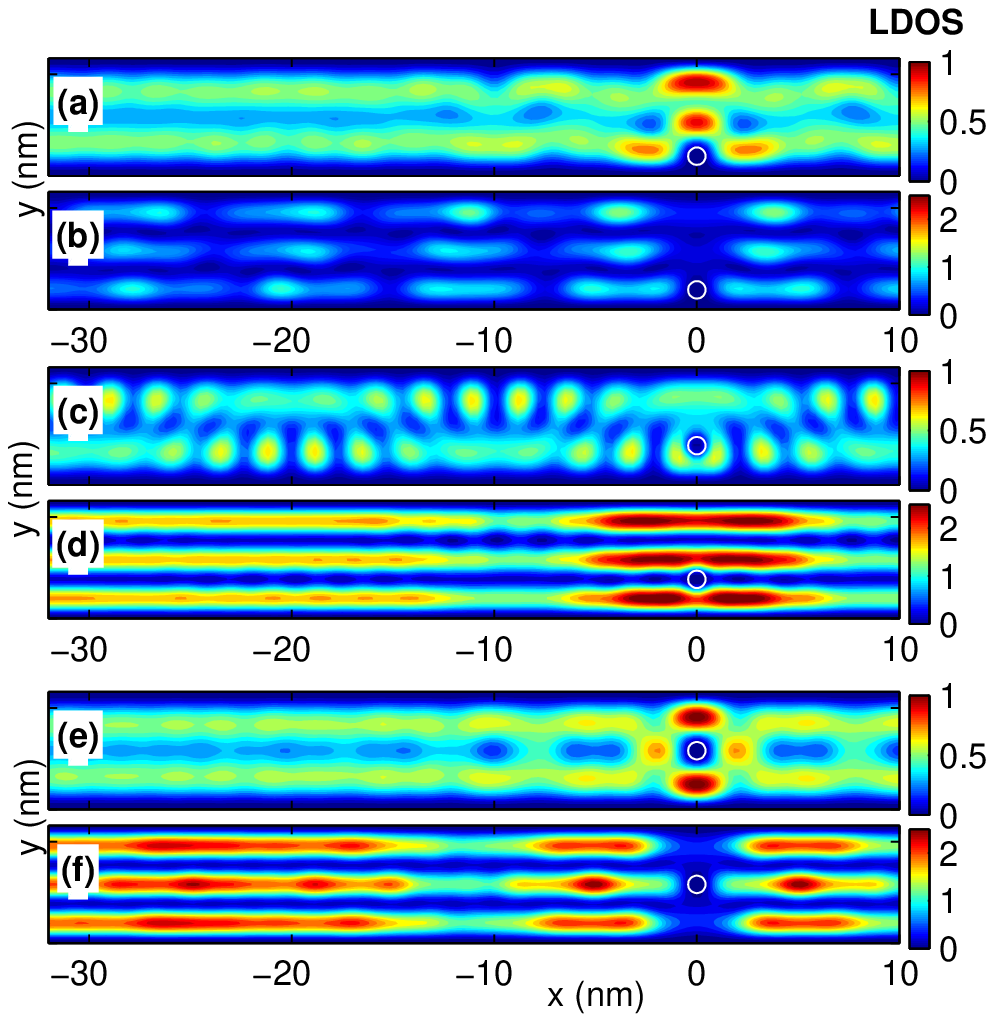}
\caption{ Contour plots of the LDOS for the resonance case with an impurity sitting at $\alpha$ (a-b), $\beta$ (c-d) and $\gamma$ (e-f).  The bias energy for panels (a,c,e) is at sub-gap $E/meV= 2.17$ and for panels (b,d,f) is at gap $E/meV=2.73$.  The open circles indicate impurities' profile where $U=0.1U_0$.} \label{sfig.3}
\end{figure*}

\begin{figure*}
\includegraphics[width=1.4\columnwidth]{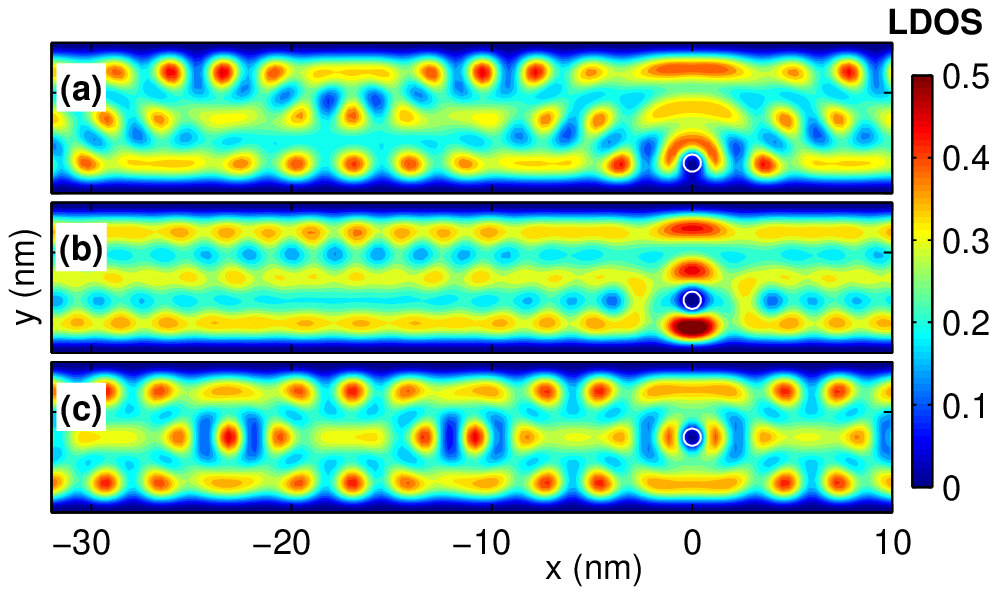}
\caption{ Contour plots of the LDOS for the off-resonance case with an impurity sitting at $\alpha'$ (a), $\beta'$ (b) and $\gamma'$ (c) at gap $E/meV=0.76$.  The open circles indicate impurities' profile where $U=0.1U_0$.} \label{sfig.4}
\end{figure*}

Next, we consider an impurity with the same parameters as in the main paper at positions $(0,\alpha)$, $(0,\beta)$ and $(0,\gamma)$ for resonance case and positions $(0,\alpha')$, $(0,\beta')$ and $(0,\gamma')$ for off-resonance case.  These are defined in Fig.~\ref{sfig.1}.  In Fig.~\ref{sfig.2}, we show the spatial distribution of the amplitude of the order parameter in the presence of the impurity.  In Fig.~\ref{sfig.3} and Fig.~\ref{sfig.4}, we show the corresponding LDOS for resonance and off-resonance cases, respectively.  All the results are similar to the ones presented in the main paper.  The effect of the impurity depends on its position and shows opposite behaviors for the resonant and off-resonant cases.  These are due to the differences in the subband energy spectrum.

\begin{figure*}
\includegraphics[width=1.4\columnwidth]{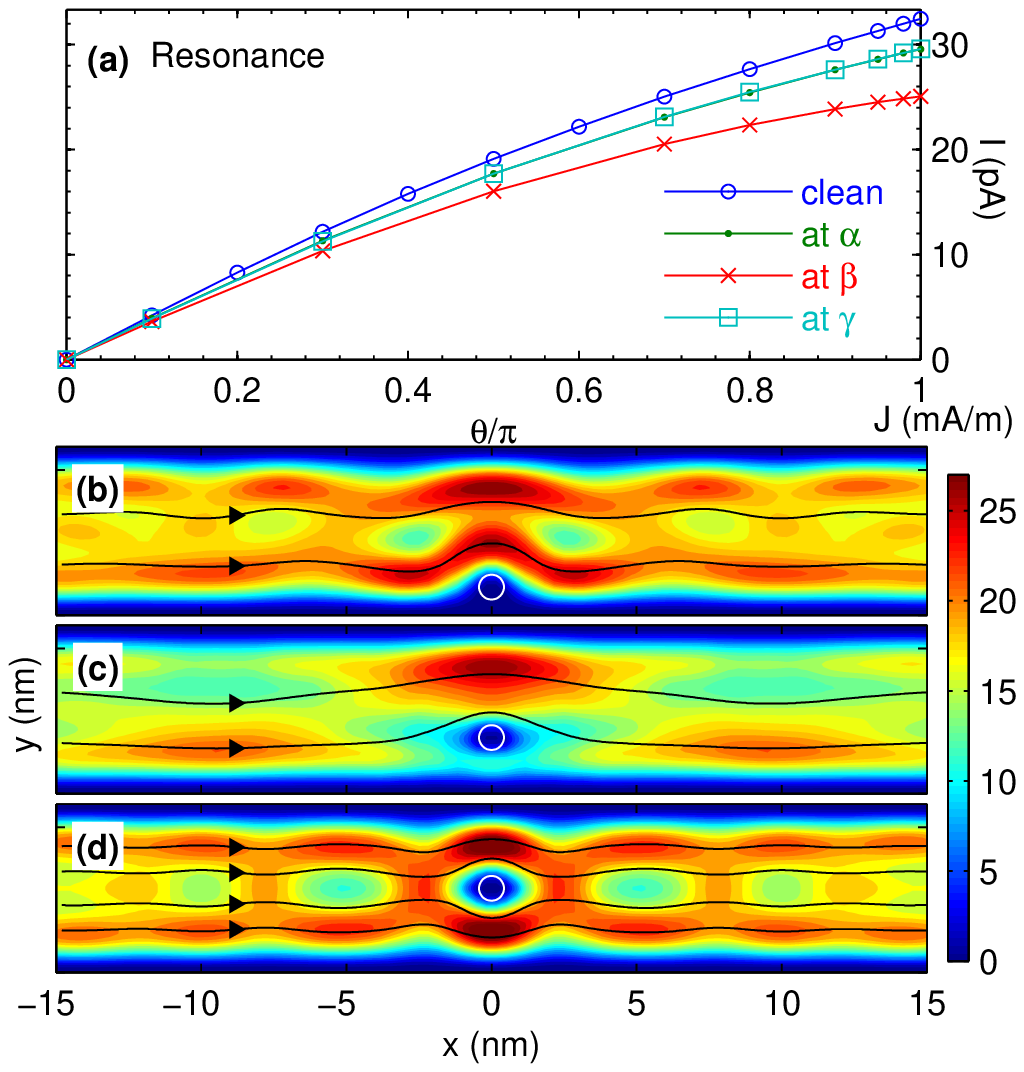}
\caption{ Josephson current for the resonance case.  (a) Current-phase relation for clean limit, with impurity at $\alpha$, $\beta$ and $\gamma$. (b)-(d) The spatial distributions of current for an impurity at $\alpha$, $\beta$ and $\gamma$ for phase differences where currents reach their critical Josephson currents.  The color indicates the amplitude of the current and the streamlines indicate the direction of the current. The open circles indicate the impurity profile for which $U=0.1U_0$.} \label{sfig.5}
\end{figure*}

\begin{figure*}
\includegraphics[width=1.4\columnwidth]{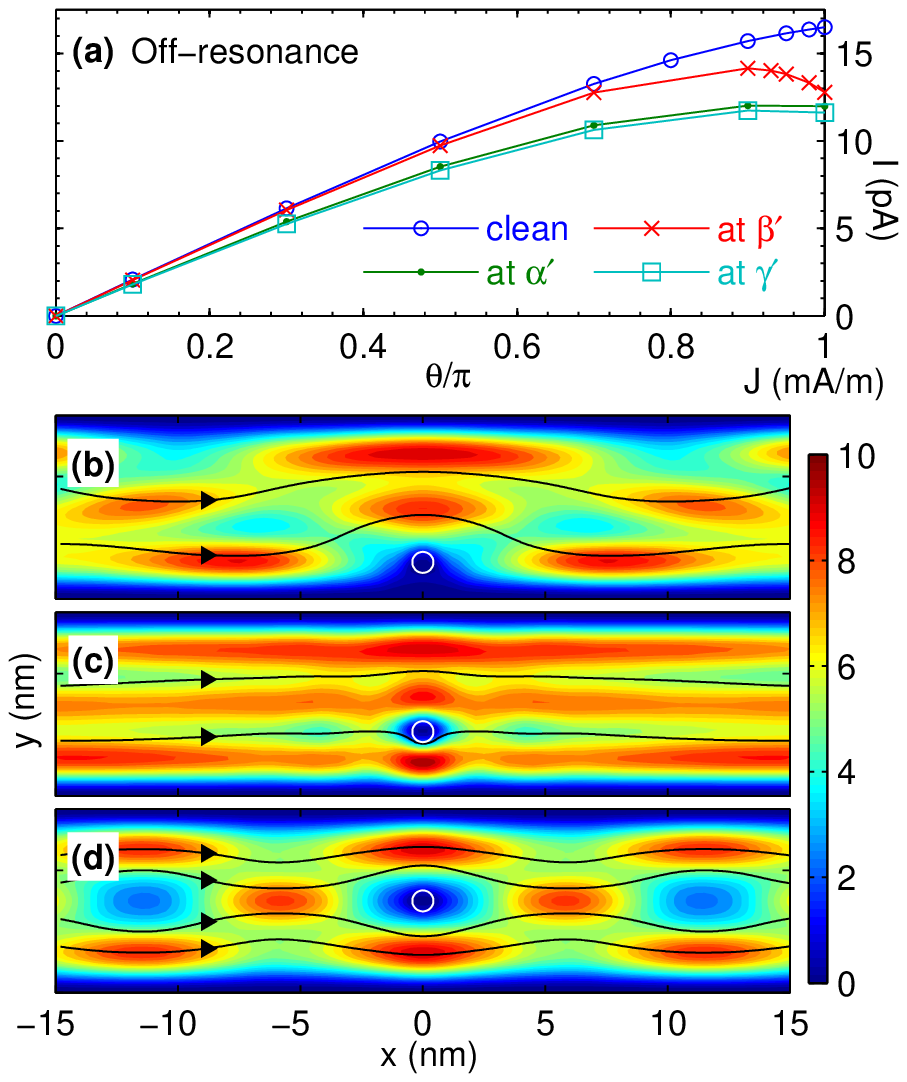}
\caption{ Josephson current for the off-resonance case.  (a) Current-phase relation for clean limit, with impurity at $\alpha'$, $\beta'$ and $\gamma'$. (b)-(d) The spatial distributions of current for an impurity at $\alpha'$, $\beta'$ and $\gamma'$ for phase differences where currents reach their critical Josephson currents.  The color indicates the amplitude of the current and the streamlines indicate the direction of the current. The open circles indicate the impurity profile for which $U=0.1U_0$.} \label{sfig.6}
\end{figure*}

Finally, we show the Josephson current for the resonance case in Fig.~\ref{sfig.5} and off-resonance case in Fig.~\ref{sfig.6}.  All other parameters are the same as in the main paper such as the length of the junction link $L_j=30nm$.  Again all results are similar to the ones shown in the main paper but the impurity has now less overall effect on the total current.  The reasons are that (1) The relative size of the impurity becomes smaller when the width of the nanowire increases.  (2) More subband quasiparticles take part in the current transport, which results in the current distribution along the lateral direction being more uniform.

In conclusion, all the additional results shown here support the explanations presented in the main paper.  Meanwhile, this shows that the effect of the impurities decrease with increase of the width of the nanowire.  Finally, as expected all the global properties converge to the bulk ones and Anderson's theorem is recovered.

\begin{thebibliography}{0}

\bibitem{revmod1}
  \Name{Balatsky A. V., Vekhter I.  \and Zhu J.-X.}
  \REVIEW {Rev. Mod. Phys.} {78}{2006}{373}.
\bibitem{probe}
  \Name{Liu B. \and Eremin I.}
  \REVIEW {Phys. Rev. B} {78}{2008}{014518}.
\bibitem{HuiHu}
  \Name{Hu H., Jiang L., Pu H., Chen Y. \and Liu X.-J.}
  \REVIEW {Phys. Rev. Lett.} {110}{2013}{020401}.
\bibitem{pin}
  \Name{Tinkham M.}
  \Book{Introduction to Superconductivity}
  \Publ{Dover Publications, Mineola, N.Y.}
  \Year{2004}.
\bibitem{disorder1}
  \Name{Fisher M. P. A.}
  \REVIEW {Phys. Rev. Lett.} {65}{1990}{923}
\bibitem{disorder2}
  \Name{Ghosal A., Randeria M. \and Trivedi N.}
  \REVIEW {Phys. Rev. Lett.} {81}{1998}{3940}.
\bibitem{disorder3}
  \Name{Dubi Y., Meir Y. \and Avishai Y.}
  \REVIEW {Nature(London)} {449}{2007}{876}.
\bibitem{disorder4}
  \Name{Bouadim K., Loh Y. L., Randeria M. \and Trivedi N.}
  \REVIEW {Nat. Phys.} {7}{2011}{884}.
\bibitem{disorder5}
  \Name{Mondal M., Kamlapure A., Chand M., Saraswat G., Kumar S., Jesudasan J., Benfatto L., Tripathi V. \and Raychaudhuri P.}
  \REVIEW {Phys. Rev. Lett.} {106}{2011}{047001}.
\bibitem{disorder6}
  \Name{Sac\'{e}p\'{e} B., Dubouchet T., Chapelier C., Sanquer M., Ovadia M., Shahar D., Feigel'man M. \and Ioffe L.}
  \REVIEW {Nat. Phys.} {7}{2011}{239}.
\bibitem{Anderson}
  \Name{Anderson P. W.}
  \REVIEW {J. Phys. Chem. Solids} {11}{1959}{26}.
\bibitem{nanoscale}
  \Name{Singh Meenakshi, Sun Yi \and Wang Jian}
  \Book{Superconductivity in Nanoscale Systems, Superconductors - Properties, Technology, and Applications}
  \Publ{InTech Publisher, Rijeka, Croatia}
  \Year{2012}.
\bibitem{slip1}
  \Name{Lau C. N., Markovic N., Bockrath M., Bezryadin A. \and Tinkham M.}
  \REVIEW {Phys. Rev. Lett.}{87}{2001}{217003}.
\bibitem{slip2}
  \Name{Michotte S., M\'{a}t\'{e}fi-Tempfli S., Piraux L., Vodolazov D. Y. \and Peeters F. M.}
  \REVIEW {Phys. Rev. B}{69}{2004}{094512}.
\bibitem{slip3}
  \Name{Li P., Wu P. M., Bomze Y., Borzenets I. V., Finkelstein G. \and Chang A. M.}
  \REVIEW {Phys. Rev. Lett.} {107}{2011}{137004}.
\bibitem{slip4}
  \Name{Astafiev O. V., Ioffe L. B., Kafanov S., Pashkin Y. A., Arutyunov K. Y., Shahar D., Cohen O. \and Tsai J. S.}
  \REVIEW {Nature(London)} {484}{2012}{355}.
\bibitem{slip5}
  \Name{Vanevi\'{c} M. \and Nazarov Y. V.}
  \REVIEW {Phys. Rev. Lett.} {108}{2012}{187002}.
\bibitem{slip6}
  \Name{Murphy A., Weinberg P., Aref T., Coskun U. C., Vakaryuk V., Levchenko A. \and Bezryadin A.}
  \REVIEW {Phys. Rev. Lett.} {110}{2013}{247001}.
\bibitem{slip7}
  \Name{Semenov A. G. \and Zaikin A. D.}
  \REVIEW {Phys. Rev. B} {88}{2013}{054505}.
\bibitem{slip8}
  \Name{Mooij J. E. \and Nazarov Y. V.}
  \REVIEW {Nat. Phys.} {2}{2006}{169}.
\bibitem{app4}
  \Name{Natarajan C. M., Tanner M. G. \and Hadfield R. H.}
  \REVIEW {Supercond. Sci. Technol.} {25}{2012}{063001}
\bibitem{app5}
  \Name{Goldtsman G. N., Okunev O., Chulkova G., Lipatov A., Semenov A., Smirnov K., Voronov B., Dzardanov A., Williams C. \and Sobolewski R.}
  \REVIEW {Appl. Phys. Lett.} {79}{2001}{705}
\bibitem{topo}
  \Name{Qi X.-L. \and Zhang S.-C.}
  \REVIEW {Rev. Mod. Phys.} {83}{2011}{1057}
\bibitem{majorana1}
  \Name{Rodrigo J. G., Crespo V., Suderow H., Vieira S. \and Guinea F.}
  \REVIEW {Phys. Rev. Lett.} {109}{2012}{237003}
\bibitem{majorana2}
  \Name{Das A. , Ronen Y., Most Y., Oreg Y., Heiblum M. \and Shtrikman H.}
  \REVIEW {Nat. Phys.} {8}{2012}{887}
\bibitem{qse1}
  \Name{Shanenko A. A., Croitoru M. D. \and Peeters F. M.}
  \REVIEW {Phys. Rev. B} {75}{2007}{014519}
\bibitem{qse2}
  \Name{Croitoru M. D., Shanenko A. A. \and Peeters F. M.}
  \REVIEW {Phys. Rev. B} {76}{2007}{024511}
\bibitem{qse3}
  \Name{Chen Y., Shanenko A. A. \and Peeters F. M.}
  \REVIEW {Phys. Rev. B} {81}{2010}{134523}
\bibitem{cascades}
  \Name{Shanenko A. A., Croitoru M. D. \and Peeters F. M.}
  \REVIEW {Phys. Rev. B} {78}{2008}{024505}
\bibitem{andreev1}
  \Name{Shanenko A. A., Croitoru M. D., Mints R. G. \and Peeters F. M.}
  \REVIEW {Phys. Rev. Lett.} {99}{2007}{067007};
  \Name{Shanenko A. A., Croitoru M. D. \and Peeters F. M.}
  \REVIEW {Phys. Rev. B} {78}{2008}{054505}.
\bibitem{zhang}
  \Name{Zhang L.-F., Covaci L., Milo\v{s}evi\'{c} M. V., Berdiyorov G. R. \and Peeters F. M.}
  \REVIEW {Phys. Rev. Lett.} {109}{2012}{107001}. \emph{ibid}.
  \REVIEW {Phys. Rev. B} {88}{2013}{144501}.
\bibitem{STM0}
  \Name{Yazdani A., Jones B. A., Lutz C. P., Crommie M. F. \and Eigler D. M.}
  \REVIEW {Science} {275}{1997}{1767}.
\bibitem{STM1}
  \Name{Cren T., Fokin D., Debontridder F., Dubost V. \and Roditchev D.}
  \REVIEW {Phys. Rev. Lett.} {102}{2009}{127005}.
\bibitem{STM2}
  \Name{Cren T., Serrier-Garcia L., Debontridder F. \and Roditchev D.}
  \REVIEW {Phys. Rev. Lett.} {107}{2011}{097202}.
\bibitem{STM3}
  \Name{Serrier-Garcia L., Cuevas J. C., Cren T., Brun C., Cherkez V., Debontridder F., Fokin D., Bergeret F. S. \and Roditchev D.}
  \REVIEW {Phys. Rev. Lett.} {110}{2013}{157003}.
\bibitem{STM4}
  \Name{Nishio T., An T., Nomura A., Miyachi K., Eguchi T., Sakata H., Lin S., Hayashi N., Nakai N., Machida M. \and Hasegawa Y.}
  \REVIEW {Phys. Rev. Lett.} {101}{2008}{167001}.
\bibitem{phase1}
  \Name{Arutyunov K. Y., Golubev D.S. \and Zaikin A.D.}
  \REVIEW {Phys. Rep.}{464}{2008}{1}.
\bibitem{phase2}
  \Name{Zaikin A. D., Golubev D. S., van Otterlo A. \and Zimanyi G. T.}
  \REVIEW {Phys. Rev. Lett.}{78}{1997}{1552}.
\bibitem{phase3}
  \Name{Khlebnikov S. \and Pryadko L. P.}
  \REVIEW {Phys. Rev. Lett.}{95}{2005}{107007}.
\bibitem{nbse1}
  \Name{Hor Y. S., Welp U., Ito Y., Xiao Z. L., Patel U., Mitchell J. F., Kwok W. K. \and Crabtree G. W.}
  \REVIEW {Appl. Phys. Lett.} {87}{2005}{142506}.
\bibitem{nbse2}
  \Name{Sekar P., Greyson E. C., Barton J. E. \and Odom T. W.}
  \REVIEW {J. Am. Chem. Soc.} {127}{2005}{2054}.
\bibitem{vortex1}
  \Name{Gygi F. \and Schluter M.}
  \REVIEW {Phys. Rev. B} {43}{1991}{7609}.
\bibitem{vortex2}
  \Name{Virtanen S. M. M. \and Salomaa M. M.}
  \REVIEW {Phys. Rev. B} {60}{1999}{14581}.
\bibitem{vortex3}
  \Name{Tanaka K., Robel I. \and Janko B.}
  \REVIEW {PNAS} {99}{2002}{5233}.
\bibitem{cases}
  {Details for other resonance ($L_y=5.8nm$) and off-resonance ($L_y=7.5nm$) cases, where there are three peaks in $|\Delta(y)|$, are shown as supplemental materials in arXiv:1401.4319.}
\bibitem{strinati}
  \Name{Spuntarelli A., Pieri P., \and Strinati G. C.}
  \REVIEW {Phys. Rep.} {488}{2010}{111}.
\bibitem{covaci2006}
  \Name{Covaci L. \and Marsiglio F.}
  \REVIEW {Phys. Rev. B} {73}{2006}{014503}.
{\bibitem{substrate2014}
  \Name{Romero-Berm\'{u}dez Aurelio and Garc\'{\i}a-Garc\'{\i}a Antonio M.}
  \REVIEW {Phys. Rev. B} {89}{2014}{064508}.}


\end{thebibliography}
\end{document}